\documentclass[12pt]{article}
\usepackage{amssymb}
\newcommand{\bea}{\begin{eqnarray}}
\newcommand{\eea}{\end{eqnarray}}
\newcommand{\be}{\begin{equation}}
\newcommand{\ee}{\end{equation}}
\newcommand{\G}{\Gamma}
\newcommand{\e}{\epsilon}
\newcommand{\C}[1]{(\ref{#1})}
\newcommand{\R}{{\mathbb R}}

\begin{document}

\begin{titlepage}

\begin{center}

\hfill hep-th/0208126 \\

\hfill EFI-02-98

\vspace{1.25in}

{\large \bf Holography and String Dynamics in
Time-dependent Backgrounds}\\

\vskip 1.25cm

 {
Akikazu Hashimoto$^{a}$\footnote[1]{e-mail: aki@ias.edu}
and
Savdeep Sethi$^{a,b}$}\footnote[2]{e-mail: sethi@theory.uchicago.edu}
%
{\vskip 0.5cm $^a$ School of Natural Sciences\\ Institute for
Advanced Study\\ Princeton, NJ 08540\\
\vskip 0.2cm $^b$ Enrico Fermi Institute\\ University of Chicago\\
Chicago, IL
60637}

\end{center}

\vskip 2 cm

\begin{abstract}
\baselineskip=18pt

\noindent We study the dynamics of D-branes in a smooth time-dependent
background. The theory on the branes is a time-dependent
non-commutative field theory. We find the metric and fluxes that
determine the dual holographic closed string theory. This provides a
concrete example of holography in a cosmological setting.

\end{abstract}

\end{titlepage}

\pagestyle{plain}
\baselineskip=19pt

In order to apply string theory to problems in cosmology, it is
essential that we learn how to formulate string theory in backgrounds
with explicit time dependence. Recently, various quotients of
Minkowski space-time by groups which acts both on space and time have
been studied as simple concrete realizations of string theory on
time-dependent backgrounds. Generic orbifolds of space-time give rise
to singularities such as closed time-like loops. These closed causal
curves tend to invalidate the usual techniques of perturbative string
theory because of large back-reaction.  One example of a space-time
orbifold which avoids these complications is the orbifold associated
to the ``null-brane'' geometry. This quotient was first considered
in~\cite{Figueroa-O'Farrill:2001nx}. String theory (and M theory) in
this particular quotient background has been studied independently
in~\cite{toappear}, and also considered recently
in~\cite{Liu:2002kb,Fabinger:2002kr}.

There are two particular distinguishing features of this quotient: it
preserves one-half of the flat space supersymmetry, and it preserves
an isometry along a light-cone direction. For these reasons,
null-branes appear to provide a promising laboratory in which to
explore issues of time-dependence in the simplest of settings.

For a given string theory background, it is often instructive to study
the dynamics of open strings by adding D-branes.  The physics of open
strings can be further simplified by taking decoupling limits which
freeze out massive open string modes and closed string excitations.
The limiting theory for the open strings is generally
non-gravitational, and can typically be formulated as a field
theory. However, it is possible to show in some cases that these
decoupled open strings have a dual description in terms of a theory of
gravity, along the lines of the AdS/CFT correspondence.

In this paper, we will show that D-branes in the background of
null-branes have an interesting decoupling limit.  The decoupled
theory turns out to be a novel kind of non-commutative Yang-Mills
theory whose non-commutativity parameter varies as a function of the
space-time coordinates.  We will further show that this theory has a
simple dual supergravity description.  The supergravity dual is also
time-dependent.  As such, this non-commutative field theory can be
thought of as a holographic description of a time-dependent string
theory background.  In this particular example, all of the time
dependence of the dual supergravity background is encoded in the time
dependence of the non-commutativity parameter on the field theory
side.  From this explicit example, we are led to the conclusion that
the holographic description of more realistic cosmological space-times
may very well require non-local field theory.

Let us begin by reviewing the geometry of the null-brane. This is
simply an orbifold of 3+1-dimensional Minkowski space-time
\be ds^2 = -2 dx^+ dx^- + dx^2 + dz^2 \ee
by the identification
\be
\begin{array}{lcl}
x^+ & \sim & x^+  \\
x  & \sim & x + 2 \pi x^+ \\
x^- & \sim & x^- + 2 \pi x + {1 \over 2} (2 \pi)^2 x^+ \\
z & \sim & z + 2 \pi R.
\end{array}
\ee

This space can also be described in terms of the coordinates
\cite{Figueroa-O'Farrill:2001nx,toappear}
\be \label{hatted}
\hat x^+ = x^+ ,\quad \hat x^- = x^- - {z x \over R} + {z^2
x^+ \over 2 R^2} ,\quad \hat x = x - {z x^+ \over R},
\quad \hat z = \frac{z}{R} \ee
in which the orbifold acts only on a single coordinate
\be
\hat z \sim \hat z + 2\pi .
\ee
The metric in these coordinates takes the form
\be \label{newmetric} ds^2 = -2 d \hat x^+ d \hat x^- + d \hat x^2 +
((\hat x^+)^2 + R^2) d \hat z^2 + 2 (\hat x^+ d \hat x - \hat x d \hat x^+)
 d \hat z . \ee

Consider embedding this geometry into type IIA string theory by
appending six flat additional directions. The quotient preserves the
half of the supersymmetries which satisfy,
\be\label{quotcond}
\G^+ \e =0.
\ee

We can now consider placing a D2-brane extended along the $x^+, x^-,$
and $x$ directions, and localized in the $z$ direction.   Such
D2-brane preserves spinors that satisfy
\be\label{branecond}
\e^+ = \G^0 \G^1 \G^2 \e^-,
\ee
where $\e^+$ and $\e^-$ are spinors with definite chirality under
 $\G^0 \G^1\cdots\G^9.$
Conditions \C{quotcond}\ and \C{branecond}\ are satisfied
simultaneously by $8$ of the $32$ supersymmetries of the type IIA
theory.\footnote{The supersymmetry of branes in the null-brane geometry
was also considered recently in \cite{Figueroa-O'Farrill:2002tb}.}

In order to extract the decoupled theory on a D2-brane, one usually
scales $\alpha'$ to zero while keeping the two-dimensional Yang-Mills
coupling $g_{YM}^2 = g_s/\sqrt{\alpha'}$ fixed.\footnote{We have
ignored numerical factors of order one.} Here, however, there will be
infinitely many copies of this D2-brane in the covering space because
of the orbifolding. Therefore, we need to account for the winding
modes which arise from the strings which stretch from a D2-brane to
its image. One way to do this is to scale $R = \alpha' /\widetilde R$
keeping $\widetilde R$ fixed.  Then, in the $\alpha' \rightarrow 0$
limit, all but the ground state of the strings in the wound sector
become infinitely massive and decouple.  The surviving wound strings
can then be thought of as momentum modes on a D3-brane which now wraps
the dual coordinate $\widetilde z$ whose radius is $\widetilde
R$.\footnote{Exchanging the M-theory circle and the T-duality circle
in the type IIA description of the above construction gives rise to
the S-dual theory in type IIB.}

We have not yet stated precisely what decoupled theory actually lives
on the D3-brane. In order to address this question, it is useful to
define a new set of coordinates by the change of
variables~\cite{Liu:2002ft},
\be
\begin{array}{lcl}
 x^+ & = & y^+ \\
x & = & y^+ y \\
x^- & = & y^- + {1 \over 2} y^+ y^2.
\end{array} \label{xz}
\ee
Unlike the transformation of \C{hatted}, this change of variables is
singular, but it has two merits. The quotient identification is
simple,
\be
\begin{array}{lcl}
 y^+ & \sim & y^+ \\
y^{\ \ } & \sim & y + 2 \pi \\
y^- & \sim & y^- \\
z & \sim & z + 2 \pi R,
\end{array}
\ee
and the metric is also simple: \be ds^2 = -2 dy^+ dy^- + (y^+)^2 dy^2
+ dz^2. \ee In particular, there are no off-diagonal terms
involving $dz$.  We observe that the D3-brane in question is
equivalent to what one obtains after the following sequence of steps:
\begin{enumerate}
\item Start with a D3-brane wrapped on $\R^{1,2} \times S^1$ with metric
\be ds^2 = -2 dy^+ dy^- + (y^+)^2 dy^2 + d{\widetilde z}^2,
\qquad \widetilde z \sim \widetilde z + 2 \pi \widetilde R. \ee
\item T-dualize along $\widetilde z$ to obtain D2-branes on
$\R^{1,2} \times S^1$  localized on the $z$ circle with metric,
\be ds^2 = -2 dy^+ dy^- + (y^+)^2 dy^2 + d z^2, \qquad
z \sim  z + 2 \pi  R, \qquad R = {\alpha' \over \widetilde R}
\label{metric}\ee
\item Now ``twist'' the $y$ coordinate with respect to $z$ so that
\be z \sim z + 2 \pi R, \qquad y \sim y + 2 \pi , \label{twist} \ee
and introduce the new coordinate:
\be \widetilde{y} = y - {z \over R}. \label{ztilde}\ee
\item Lastly, T-dualize back along the $z$ direction.
\end{enumerate}
The sequence of steps enumerated above is a familiar one.  Had the
metric on $\R^{1,2} \times S^1$ been the flat metric,
\be ds^2 = - dt^2 + dx^2 + dy^2 + dz^2, \ee
and had we twisted the $y$ coordinate by the identification
\be z \sim z + 2 \pi R, \qquad y \sim y + {\Delta^2 \over \alpha'} R, \ee
in following these steps, we would have obtained the familiar
non-commutative Yang-Mills theory with~\cite{Douglas:1998fm}
\be \theta^{yz} = -\theta^{z y} = \Delta^2  .  \ee
A similar procedure, where one twists the plane transverse to the
brane, was used to construct non-commutative dipole theories where the
dipole length is proportional to the
$R$-charge~\cite{Bergman:2001rw,Dasgupta:2001zu}.

In analogy, it is natural to propose that the theory on the D3-brane
obtained by T-dualizing a D2-brane in the null-brane background is a
non-commutative Yang-Mills theory with a $*$-product defined by
\be  f * g =  
f g
+ i \theta^{\widetilde y \widetilde z} \left(
\partial_{\widetilde y} f\,  \partial_{\widetilde z} g 
- 
\partial_{\widetilde z} f\,  \partial_{\widetilde y} g  \right) + {\cal O}(\theta^2). 
\label{yzstar}\ee
Since we shifted $y$ by a number of order one,
\be 1 = {\theta^{\tilde y \tilde z} \over \alpha' }R
= {\theta^{\tilde y \tilde z}  \over \widetilde R}\,  \ee
so $\theta^{\tilde y \tilde z} = \widetilde R$.

Since the $y$ coordinates are somewhat singular compared to the $x$
coordinates, let us recast this product in terms of the original
coordinates. This can be done straightforwardly. Simply map from $y$
coordinates to the $x$ coordinates using the relation \C{xz} with
$y$ replaced by $\widetilde y$ so that one finds\footnote{This
procedure of mapping $x$ to $y$, replacing $y$ by $\widetilde y$, and
mapping back to $x$ is what gives rise to the $\hat x$ coordinates
used in \C{hatted}.}
\be {\partial \over \partial \widetilde y} = x^+ {\partial \over \partial
  x}
+ x {\partial \over \partial x^-} . \ee
One can therefore write the $*$-product in the $x$ coordinates in the
standard form
\be f(x)*g(x) = f(x) g(x) + 
i \theta^{\mu \nu} \, {\partial_\mu} f(x) \, {\partial_\nu} g(x)  
+ {\cal O}(\theta^2) \ee
with
\be\theta^{xz} = -\theta^{z x} = \widetilde R  x^+, \qquad
\theta^{x^-z} = -\theta^{z x^-} = \widetilde R x\label{theta},\ee
and with all the other components of $\theta^{\mu \nu}$ vanishing.  It
is easy to verify that $\theta^{\mu \nu}$ obeys the identity
\be \theta^{il} \partial_l \theta^{jk} +  \theta^{jl}
\partial_l \theta^{ki}  +  \theta^{kl} \partial_l \theta^{ij}  = 0\ee
which ensures that the $*$-product is associative and that all of the
terms higher order in $\theta$ can be generated systematically using
the methods of \cite{Kontsevich:1997vb,
Cattaneo:1999fm,Cornalba:2001sm}.  What we have, therefore, is a
concrete string theory realization of a non-commutative Yang-Mills
theory whose non-commutativity parameter varies over space-time.

In the remainder of this note, we will provide two additional checks
that this non-commutative Yang-Mills theory is the decoupled theory of
open strings on the brane: $(1)$ the analysis of open string variables
in the brane probe approximation, and $(2)$ the analysis of the dual
supergravity solution in the sense of AdS/CFT correspondence.

To see what background the D3-brane sees, let us find the T-dual of
the metric (\ref{metric}) with the twist (\ref{twist})
explicitly. Because of the twist, it is useful to use the coordinate
$\widetilde y$ introduced in (\ref{ztilde}). In terms of $\widetilde
y$ and $z$, the metric (\ref{metric}) takes the form
\be ds^2 = -2dy^+ dy^- + {R^2 (y^+)^2 \over R^2 + (y^+)^2} d \widetilde y^2
+ {R^2 + (y^+)^2 \over R^2}\left(dz + {R (y^+)^2 \over R^2 + (y^+)^2}
d \widetilde y\right)^2. \ee
Using the formula for T-duality which can be found, for example, in
\cite{Giveon:1994fu}, we find that the dual background is given by
\bea
ds^2 &=& -2dy^+ dy^- + {R^2 \over R^2 + (y^+)^2} \left( (y^+)^2 d \widetilde
y^2  + d \widetilde z^2 \right) \cr
B &=&  {R (y^+)^2 \over R^2 + (y^+)^2} \,  d \widetilde y
\wedge d\widetilde z \label{IIbBG} \cr
e^{\phi} & = & g_s \sqrt{R^2 + (y^+)^2 \over R^2}.
\eea
The metric, $B$-field, and coupling explicitly depend on
$y^+$. However, these are the parameters natural for the closed
strings which need not coincide with the parameters natural for the
open strings, especially in the presence of a background $B$-field.
The map between open and closed string parameters was derived in
\cite{Seiberg:1999vs} for the case where the metric and $B$-fields do
not vary.  Although this is not the case here, let us simply apply the
formulae of \cite{Seiberg:1999vs} by assuming that the background is
sufficiently locally constant. The open string metric, $G$, and the
closed string metric, $g$, are related by
\be G^{\mu \nu} + i {\theta^{\mu \nu} \over \alpha'}
= (g_{\mu \nu} + i B_{\mu \nu})^{-1}.  \ee
Substituting the values of $g_{\tilde y \tilde y}$, $g_{\tilde z
\tilde z}$ and $B_{\tilde y \tilde z}$ read off from \C{IIbBG}, we
find
\be G^{\tilde y \tilde y} = (y^+)^{-2},
\qquad G^{\tilde z \tilde z} = 1,
\qquad \theta^{\tilde y \tilde z}
= \widetilde R. \ee
This is in complete agreement with the form of the $*$-product
anticipated in \C{yzstar}.  Along similar lines, it is easy to show
that the open string coupling
\be G_s = e^{\phi} \sqrt{ {\det(g+B) \over \det g}} = g_s \ee
is non-varying, and sets the value of the Yang-Mills coupling constant:
\be g_{YM}^2 = G_s. \ee
We therefore conclude that although both the background metric and
the coupling vary over space-time in the closed string variables,
the NS-NS $B$-field also varies precisely in a manner that makes
the open string metric and the coupling static.\footnote{For
theories with constant non-commutativity, this phenomena has been
argued to be quite generic~\cite{Berman:2000jw}.} Therefore, we
end up with a non-commutative Yang-Mills theory defined on a flat
static background, with all of the space-time dependence absorbed
into the variation of the non-commutativity length scale.

The non-commutative field theory described above is a decoupling limit
of open strings on a D-brane.  It is therefore natural to ask if a
dual supergravity description can be found by taking the near horizon
limit of the supergravity solution which describes this
brane. Fortunately, because the null-brane is a simple orbifold of
flat space, it turns out to be possible to construct the supergravity
dual explicitly.

The near horizon geometry we seek can be constructed by following the
sequence of steps involving a T-duality and a twist described earlier.
In fact, the construction closely mimics the one used in
\cite{Bergman:2001rw,Alishahiha:2002ex} to construct the supergravity
dual of the dipole theory.

\begin{enumerate}
\item Start with the metric of the flat D3-brane but using the $y$
coordinates
\be ds^2 = f^{-1/2}(-2 dy^+ dy^- + (y^+)^2 dy^2 + d \widetilde z^2)
+ f^{1/2}(dr^2 + r^2 d \Omega_5^2) \ee
where $f$ is the harmonic function of the D3-brane
\be f = 1 + {g N \alpha'^2  \over r^4} . \ee
The $\widetilde z$ coordinate is compactified on a circle of radius
$\widetilde R$. We will concentrate on the components of the metric
parallel to the brane since the transverse components are unaffected
in the following discussion.

\item T-dualize along $\widetilde z$. We now have the type IIA metric for
D2-branes smeared along $z$,
\be ds^2 = f^{-1/2}(-2 dy^+ dy^- + (y^+)^2 dy^2) + f^{1/2} dz^2. \ee

\item Twist the $y$ coordinate so that the identification under the
shift in $z$ becomes
\be z \sim z + 2 \pi R, \qquad y \sim y + 2 \pi  .\ee
Introduce a new coordinate,
\be \widetilde y = y - {z \over R}, \ee
so that the identification of $\widetilde y$ is trivial
\be z \sim z + 2 \pi R, \qquad \widetilde y \sim \widetilde y.\ee
The metric in terms of these new variables then becomes
\be ds^2 = -2 f^{-1/2}dy^+ dy^- + {f^{1/2} R^2 (y^+)^2 \over f R^2 +
(y^+)^2} d \widetilde y^2 + {fR^2 + (y^+)^2 \over f^{1/2} R^2}\left(dz +
{R(y^+)^2 \over f R^2 + (y^+)^2}d \widetilde y\right)^2. \ee

\item T-dualize back along the $z$ direction. Applying formulae that
can be found in \cite{Giveon:1994fu} gives
\bea ds^2 &=& f^{-1/2} \left(-2dy^+dy^- +
{f R^2  \over f R^2 +
(y^+)^2} \left( (y^+)^2 d \widetilde y^2  + d \widetilde z^2 \right)
\right) \label{d3sugra}\\
B & = & {R (y^+)^2 \over f R^2 + (y^+)^2} d\widetilde y \wedge d
\widetilde z\
\eea
for the metric and $B$-field.  The dilaton and the background
Ramond-Ramond field strengths can also be found by following this
duality.
\end{enumerate}

Now that we have found the supergravity solution \C{d3sugra}\
describing the D3-brane in the background \C{IIbBG}, we can apply the
standard near horizon scaling limit
\be \alpha' \rightarrow 0, \qquad
U = {r \over \alpha'} =\mbox{fixed},
\qquad \widetilde R = {\alpha' \over R} = \mbox{fixed} \ee
to derive the supergravity dual of the decoupled theory on the
brane. The metric in string frame is found to be
\be ds^2 = \alpha' \left( {U^2 \over \sqrt{\lambda}} \left( -2 dy^+ dy^-
+ {(y^+)^2 d\widetilde y^2 + d\widetilde z^2
\over 1 + {\widetilde{R}^2 (y^+)^2 U^4 \over
\lambda}} \right) + {\sqrt{\lambda} \over U^2} ( dU^2 + U^2 d
\Omega_5^2) \right), \label{nhdual} \ee
where $\lambda  = g_{YM}^2 N$ is the 't Hooft coupling.

Note that the form of the supergravity solution very closely resembles
the supergravity dual of non-commutative Yang-Mills theory described
in~\cite{Hashimoto:1999ut,Maldacena:1999mh}
\be ds^2 = \alpha' \left( {U^2 \over \sqrt{\lambda}} \left(-dt^2 + dx^2 +
{ d z^2 + d y^2 \over 1 + {\Delta^4 U^4 \over
\lambda}} \right) + {\sqrt{\lambda} \over U^2} ( dU^2 + U^2 d
\Omega_5^2) \right), \label{ncymsg} \ee
where $\Delta^2 = \theta^{yz}$ is the non-commutativity parameter.
By comparing the form of (\ref{nhdual}) and (\ref{ncymsg}), we can
identify the scale of non-commutativity in the $(\widetilde y,\widetilde
z)$-plane to be $\widetilde R$, which is consistent with the form of the
$*$-product in (\ref{yzstar}).

The main claim of this paper is the duality between a non-commutative
Yang-Mills theory with non-commutativity parameter (\ref{theta}) and
string theory on the background (\ref{nhdual}).  The background
(\ref{nhdual}) is the T-dual of the null-brane geometry with large
gravitational backreaction due to the presence of the D3-branes.  This
background inherits the explicit time-dependence of the null-brane
geometry. The non-commutative field theory can therefore be
interpreted as a holographic description of a time-dependent string
background.  It is interesting that although various fields, including
the metric, dilaton, and the $B$-field, vary over space-time in the
bulk description, the only space-time varying parameter on the field
theory side appears to be the non-commutativity parameter.

The decoupled field theory is an interesting theory in its own right.
To our knowledge, this is the first concrete string theory realization
of a non-commutative field theory with non-constant non-commutativity
parameter.  Since the non-commutativity parameter $\theta^{\mu \nu}$
does not have an $x^+$ component, the action will not contain higher
$x^+$ derivative terms, so it would be natural to quantize the theory
treating $x^+$ as time. It would, nonetheless, be non-trivial to
quantize this theory as the action depends explicitly on $x^+$ through
the non-commutativity parameter. It would be interesting to explore
the standard issues of non-commutative field theory for this kind of
model, including thermodynamics, UV/IR mixing, soliton dynamics, the
gauge invariant observables, and S-duality to name just a few.

The main appeal of the duality presented in this paper is that it
establishes concretely the possibility of defining a cosmological
background using holography. It is very interesting to speculate on
the possibility of finding a holographic dual to more realistic
cosmologies along similar lines.  The example considered in this paper
clearly demonstrates that the holographic theory, should one exist,
need not be a local quantum field theory.

\section*{Acknowledgements}

The work of A.~H. is supported in part by DOE Grant No.
DE-FG02-90ER40542, and the Marvin L.~Goldberger fellowship. The work
of S.~S. is supported in part by NSF CAREER Grant No.  PHY-0094328,
and by the Alfred P. Sloan Foundation.  We would also like to
thank the Aspen Center for Physics where this work was completed.

\bibliographystyle{utphys}
\bibliography{myrefs}

\end{document}